\documentclass[12pt]{article}
\usepackage[utf8]{inputenc}
\usepackage{textcomp}
\usepackage{graphicx}
\usepackage{amsmath}
\usepackage{array}
\usepackage{multirow}
\usepackage{booktabs}
\usepackage{xspace}
\usepackage[hidelinks]{hyperref}

\DeclareFontEncoding{LGR}{}{}
\DeclareFontSubstitution{LGR}{cmr}{m}{n}
\DeclareTextSymbol{\textgamma}{LGR}{103}
\DeclareTextSymbolDefault{\textgamma}{LGR}

\newcommand{\mybar}[1]{\smash{$\bar{\text{#1}}$}}
\newcommand{\tbar}{\mybar{t}}
\newcommand{\ttbar}{t\tbar}
\newcommand{\mygamma}{\textgamma\xspace}

\newcommand{\TeV}{\,TeV\xspace}
\newcommand{\sqrts}[1][13]{$\sqrt{s}=#1$\TeV\xspace}

\newcommand{\ttG}{$t\bar{t}\gamma$\xspace}
\newcommand{\ctZ}{$c_{tZ}$\xspace}
\newcommand{\ctZI}{$c_{tZ}^{\mathrm{I}}$\xspace}

\begin{document}
\begin{titlepage}\pagenumbering{Alph}
\rightline{\begin{tabular}{l}
    CMS CR-2021/293 \\ 
    December 23, 2021 
\end{tabular}}

\vfill
\begin{center}\Large 
  Inclusive and differential $t\bar{t}\gamma$ measurement in the dilepton channel and effective field theory interpretation
\end{center}
\vfill
\begin{center}
    \href{mailto:gianny.mestdach@cern.ch}{\textsc{Gianny Mestdach}} \\
    \textit{Ghent University, Ghent, Belgium}
\end{center}
\begin{center}
    \textsc{on behalf of the CMS Collaboration}
\end{center}
\vfill
\begin{quotation} 

  The production cross section of $t\bar{t}\gamma$ is measured in the dilepton channel using 138 fb$^{-1}$ of proton-proton collision data recorded by the CMS experiment at $\sqrt{s}= 13 $~TeV  during
  the 2016-2018 data-taking period of the CERN LHC.
  A fiducial phase space is defined at the particle level in which both inclusive and differential results are provided.
  The $t\bar{t}\gamma$ process is sensitive to EFT operators that modify the top-photon coupling, so that tight constraints on the $c_{tZ}$ and $c_{tZ}^{\mathrm{I}}$ Wilson coefficients can be extracted.
  Finally, the EFT interpretation is repeated on a combination of this result and a measurement of $t\bar{t}\gamma$ production in the lepton+jets channel.

\end{quotation}
\vfill
\begin{quotation}\begin{center}
    PRESENTED AT
\end{center}\bigskip\begin{center}\large
    $14^\mathrm{th}$ International Workshop on Top Quark Physics\\
    online, September 13--17, 2021
\end{center}\end{quotation}
\vfill
\end{titlepage}
\def\thefootnote{\fnsymbol{footnote}}
\setcounter{footnote}{0}
\pagenumbering{arabic}

\section{Introduction}

\begin{figure}[!ht]
  \centering
  \includegraphics[width=0.3\textwidth]{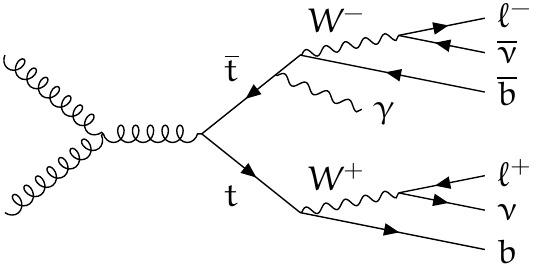}%
  \hfill%
  \includegraphics[width=0.3\textwidth]{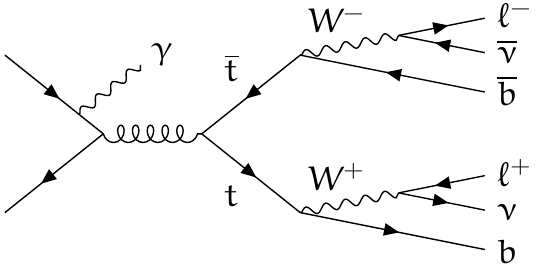}%
  \hfill%
  \includegraphics[width=0.3\textwidth]{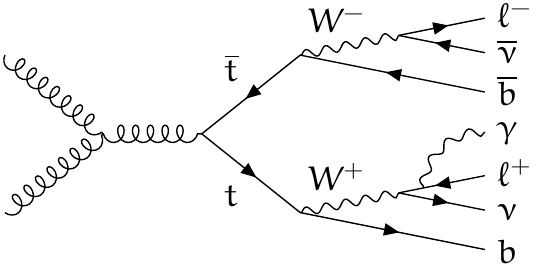}%
  \caption{%
      Leading-order Feynman diagrams for \ttG production with two leptons in the final state, where the photon is radiated by a top quark (left), by an incoming quark (middle), and by one of the charged decay products of a top quark (right).
  }
  \label{fig:feynman}
  \end{figure}

Inclusive and differential measurements of \ttG production in the dilepton channel are presented~\cite{CMS-ttGdi}.
These results were obtained using 138 fb$^{-1}$ of proton-proton collision data recorded by the CMS experiment~\cite{CMS-Experiment} at \sqrts during the 2016–2018 data-taking period of the CERN LHC. 
The signal process is defined at the particle level, and includes prompt photons radiated anywhere from initial state radiation, to the final state particles originating from the decay of the top quarks.
Representative diagrams for \ttG events where both top quarks decay into leptons are shown in Fig.~\ref{fig:feynman}.

As \ttG production is sensitive to the top-photon coupling, this measurement can be used to put constraints on modifications of this coupling.
This is done in a model independent way in the framework of standard model effective field theory (SMEFT), and tight constaints on the \ctZ and \ctZI Wilson coefficients are obtained.
Additionally, the SMEFT interpratation is repeated on a combination of the results of this measurement, and those of the measurement of \ttG in the lepton+jets channel by CMS~\cite{CMS-ttGsl}.

\section{Event selection}

Events are selected with two opposite-charge leptons, identified using a multivariate discriminator optimized to reject nonprompt leptons~\cite{CMS-tZq}.
The invariant mass of the two-lepton system $m(ll)$ is required to be $>20$ GeV and to fall outside of a window around the Z boson mass given by $|m(ll)-m_{Z}|>15$ GeV.
A similar requirement is imposed on the invariant mass of the two leptons and the photon, namely $|m(ll\gamma)-m_{Z}|>15$ GeV. These last two cuts significantly reduce the contribution from the $DY$ and $Z\gamma$ background processes, and are only applied in the $ee$ and $\mu \mu$ channels.
Photons are identified using cut-based identification (ID) criteria, and events are required to have exactly one photon in the barrel region of the detector, with no additional photons passing a looser version of the ID.
In simulated events, selected photons are categorized as prompt or nonprompt based on a matching procedure to the nearest generator particle with similar energy, and it is $t\bar{t}$ events decaying to two leptons with one prompt photon that make up the signal.

\section{Background estimation}
The backgrounds relevant to this measurement can be divided into those contributing with a prompt photon, and those that enter by a nonprompt photon being selected.
Any process can contribute to the nonprompt photon background, and this category is estimated from data using an ABCD method for which a nonprompt photon enriched sideband is created by relaxing and inverting some of the photon ID criteria.
The production of $Z \gamma$ forms the largest fraction of the prompt photon background in the $ee$ and $\mu \mu$ channels, while its contribution is negligible in the $e \mu$ channel. 
This contribution is predicted using Monte Carlo (MC) simulation, but corrections to its yield and $N_{j}/N_{b}$ distribution are derived in a control region obtained by inverting the cut applied to $m(ll\gamma)$ in the signal selection.
The other, less prominent backgrounds with prompt photons are divided into Single-t+$\gamma$ for single top processes, and Other+$\gamma$ for any other minor backgrounds (e.g. diboson production). 
These two categories are also predicted using MC simulation.

\section{Results}
\begin{figure}[!htp]
  \centering
  \includegraphics[width=0.49\textwidth]{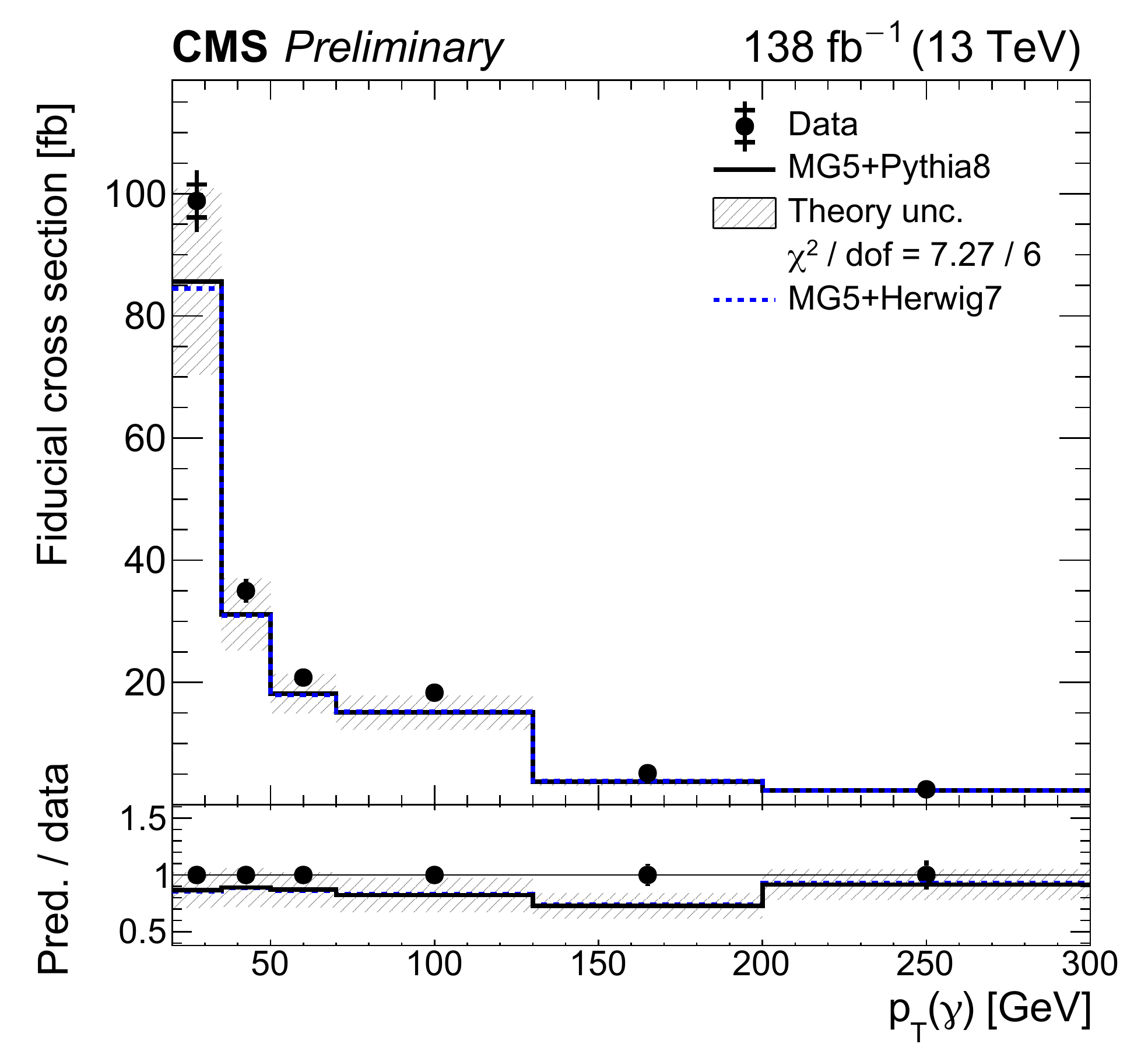}
  \includegraphics[width=0.49\textwidth]{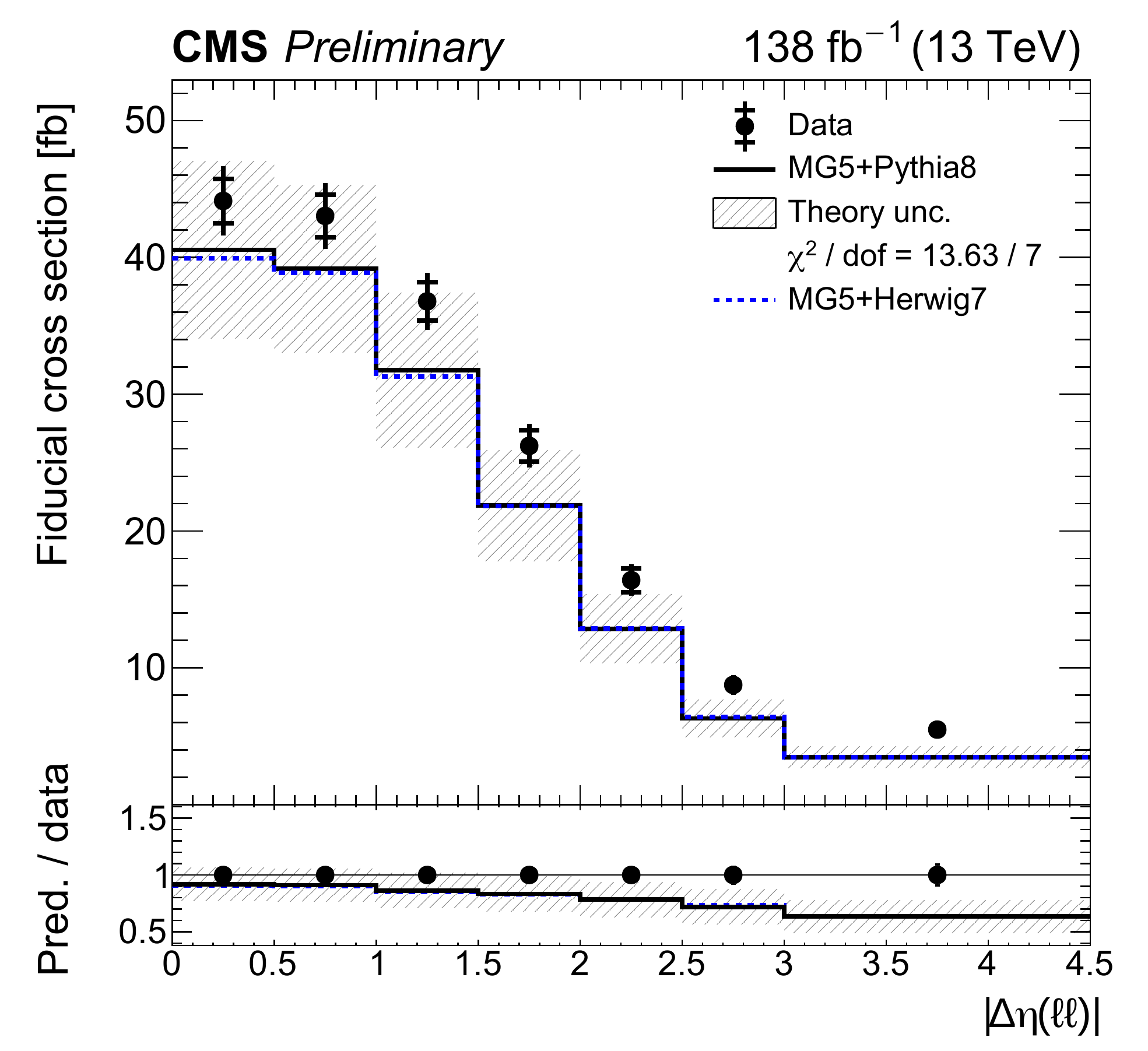}
  \caption{
  Absolute differential \ttG production cross sections as functions of $p_{\mathrm{T}}(\gamma)$ (left) and $|\Delta \eta (ll)|$ (right).
  The data are represented by points, with inner (outer) vertical bars indicating the statistical (total) uncertainties. 
  The theoretical uncertainties in the prediction using PYTHIA8 are indicated by shaded bands.}
  \label{fig:dif}
  \end{figure}

An inclusive differential cross section is obtained within the fiducial phase space using a maximum likelihood fit to the photon $p_{\mathrm{T}}$ distribution, with a value of
\begin{equation}
  174.4\pm 2.5 \textrm{ (stat)} \pm 6.1 \textrm{ (syst)} \textrm{ fb}
\end{equation}
This value is compatible with the next-to-leading order standard model prediction of $153 \pm 27 \textrm{ fb}$ within the uncertainties.
In this fit the various sources of systematic uncertainties are implemented as nuisance parameters, which can modify both the shape and yield of the predicted distributions.
Several differential results are obtained for variables ranging from photon, lepton, and jet kinematics, to the angles between these objects. 
Fig.~\ref{fig:dif} shows two of these results, 
namely for the transverse momentum of the photon and the pseudorapidity difference between the two leptons.
The obtained distributions are compared to two theory predictions obtained using different parton shower models, and the paper includes both absolute and normalized differential results. 
These differential results are obtained using an unregularized unfolding procedure for which the \verb|TUnfold| package is used. 
The response matrices have a binning that is twice as fine at the generator level than at the reconstruction level, and the binning itself is optimized for each variable in order to minimize bin-to-bin migrations.

\section{Effective field theory interpretation}
\begin{figure}[!htp]
  \centering
  \includegraphics[width=0.485\textwidth]{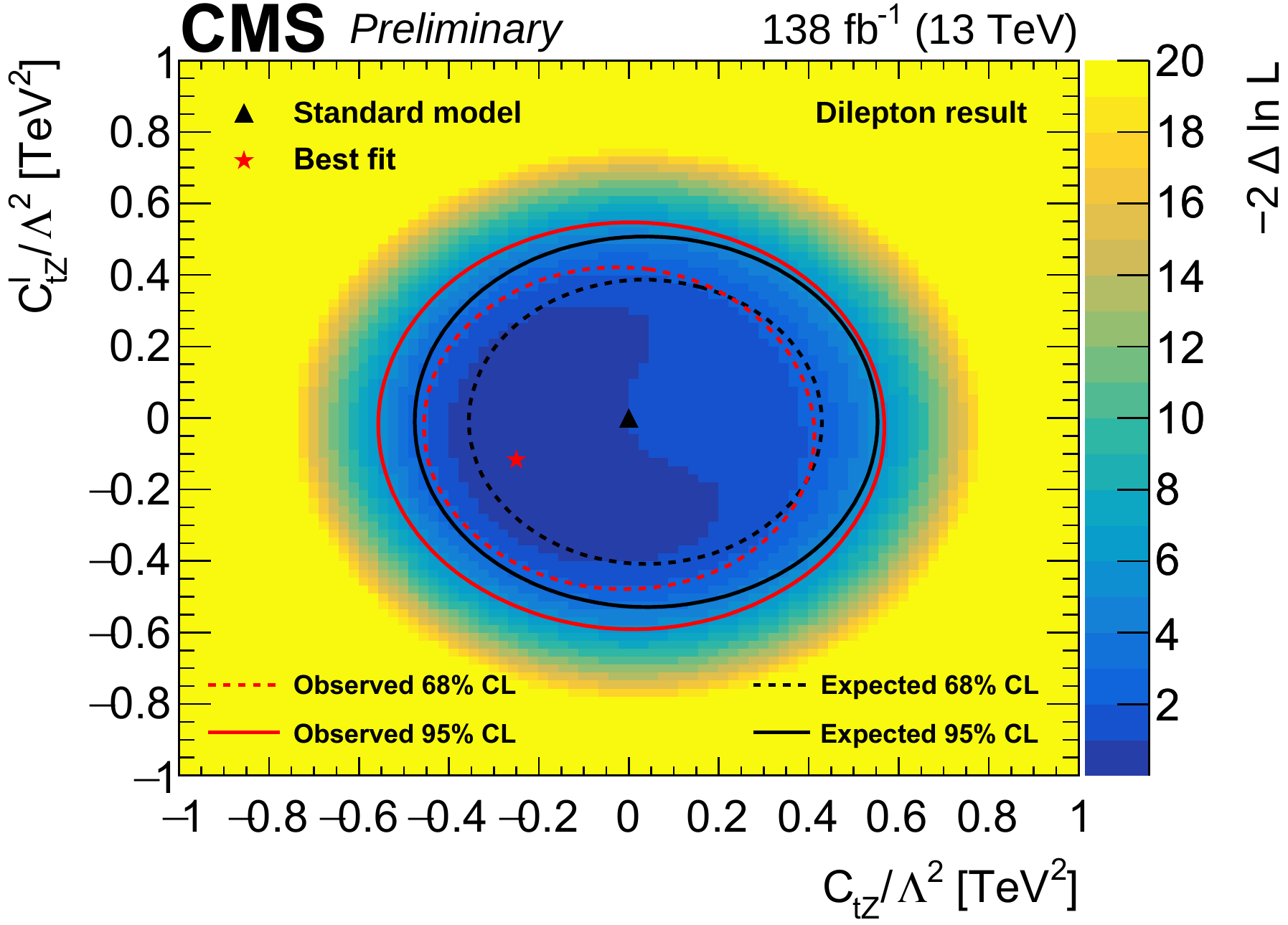}%
  \hspace{0.02\textwidth}%
  \includegraphics[width=0.485\textwidth]{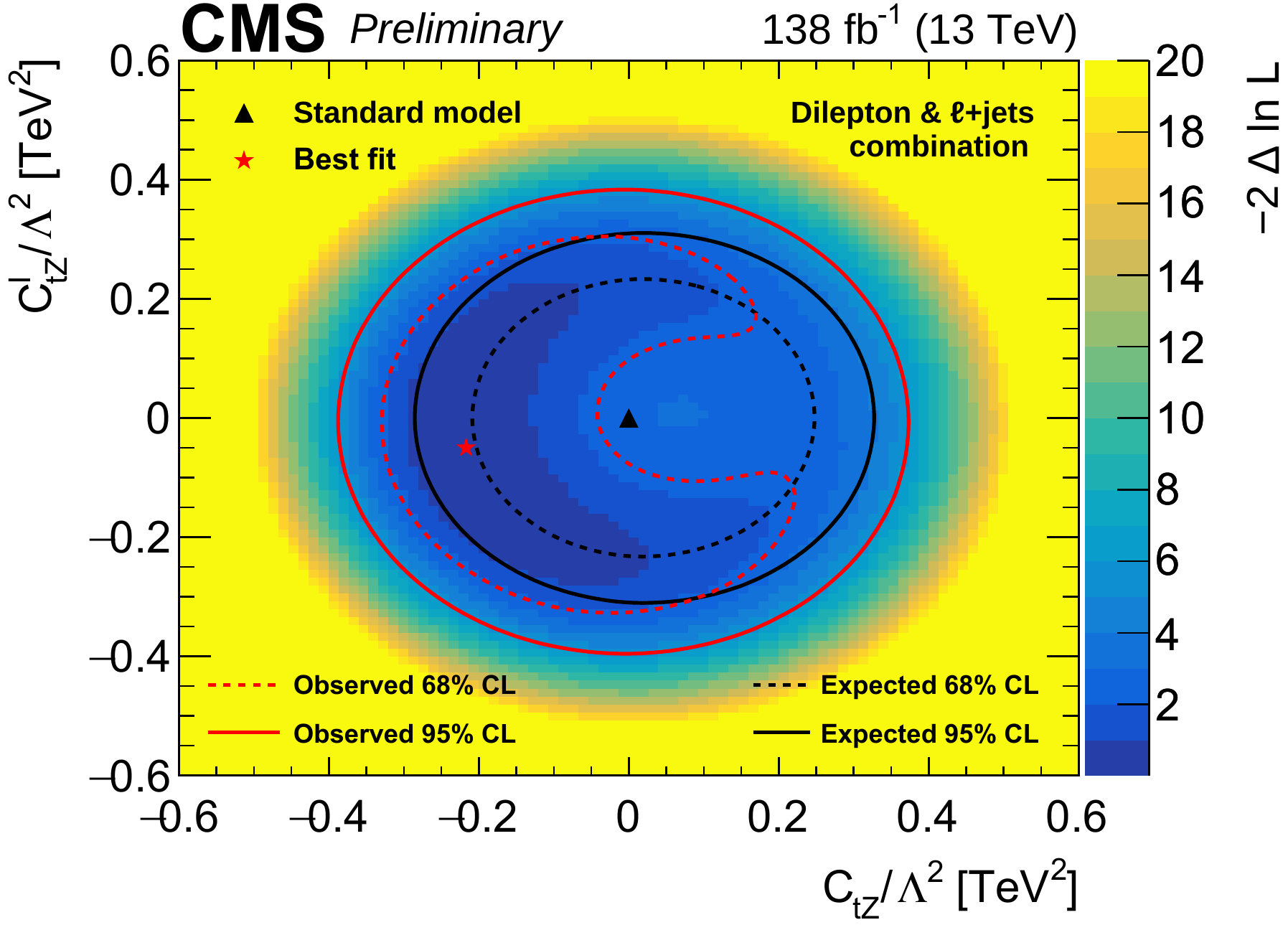}%
  \caption{%
      Result from the two-dimensional scan of the Wilson coefficients \ctZ and \ctZI using the photon $p_{\mathrm{T}}$ distribution from this analysis (left) or the combination of this analysis with the lepton+jets analysis.
  }
  \label{fig:eft2D}
  \end{figure}

From the measurement of \ttG production constraints can be put on SMEFT operators that could modify electroweak dipole moments of the top quark.
We parametrize the shape and yield of the signal as a function of \ctZ and \ctZI, of which the corresponding operators are linear combinations of the 
$O^{33}_{uB}$ and $O^{33}_{uW}$ operators as given in the Warsaw basis. The definition used in these results involves setting $c^{33}_{uW}=0$ given that the Wtb vertex
can be better probed in measurements of W helicity fractions. 
The hypothetical contributions of these operators are suppresed by a fixed \mbox{$\Lambda=1$ TeV} parameter, which means that any effects are expected to appear in the high energy tail of the $p_{\mathrm{T}}(\gamma)$ distribution.
The parametrization is in practice performed by reweighting the nominal simulation after the full analysis selection using weights obtained by taking the ratio of simulated distibutions with nonzero values for the Wilson coefficients to the nominal case at the particle level.
Fits to the photon $p_{\mathrm{T}}$ distribution similar to what is used for the extraction of the inclusive cross section are the performed as a function of the Wilson coefficients. 
This can be done both for operators individually (setting the other Wilson coeffient to zero), or for both coefficients simultaneously. 
The result of the latter is given in the left plot of Fig.~\ref{fig:eft2D}.
The same approach and operators are used in the lepton+jets analysis~\cite{CMS-ttGsl}, so that a combination of these two results can be readily performed. 
Overlap between the events in the two analyses is found to be minimal, and systematic uncertainties are correlated where necessary.
Given that the lepton+jets analysis benefits from significantly higher statistics in the high $p_{\mathrm{T}}(\gamma)$ tail, better constrains are obtained than using dilepton information.
Nevertheless, the most stringent constraints are obtained in the combination of the two, for which the 2D likelihood scan is shown in the right plot of Fig.~\ref{fig:eft2D}.

%

\end{document}